# Evidence for Newly Synthesized Lithium in Interstellar Space

D.C. Knauth*, S.R. Federman*, David L. Lambert†, P. Crane‡

*Department of Physics & Astronomy, University of Toledo, Toledo, OH 43606, USA

†Department of Astronomy, University of Texas, Austin, TX 78712, USA

‡Department of Physics & Astronomy, Dartmouth College, Hanover, New Hampshire 03755, USA and NASA Headquarters, Washington, DC 20546 USA

Astronomical observations of elemental and isotopic abundances in a variety of environments provide the means to extract the source and production route for elements as a function of time since our Galaxy formed about 15 billion years ago. Spallation reactions, where Galactic cosmic rays break apart interstellar nuclei, play an important role in the synthesis of the light elements – lithium, beryllium, and boron [1]. Measurements of the interstellar lithium isotope ratio offer a test of the proposed models of lithium synthesis in particular and light element synthesis in general [2–5]. Here we report the presence of variations in this ratio in interstellar cloud(s) toward stars associated with a site of massive star formation. Our observations reveal a sight line (o Persei) with $^7$Li/$^6$Li ratios of approximately 2, rather than the Solar System ratio of about 12.3 [6,7]. While our measured value of approximately 2 provides the first clear evidence for undiluted products of cosmic ray spallation in interstellar space and confirms a basic tenet of models of light element synthesis [2–5], the expected enhancement in elemental lithium abundance is not seen.

The observations were acquired with the Harlan J. Smith 2.7 meter telescope and the "2dcoudé" spectrograph at McDonald Observatory [8]. We observed $\zeta$ Per (B1 Ib; $V = 2.85$; $v\ sini = 59$ km s$^{-1}$) in 1996 November and 1997 January and o Per (B1 III; $V = 3.83$; $v\ sini = 85$ km s$^{-1}$) in 1998 January. High resolution spectra (R $\sim$ 170,000) in two spectral regions were obtained, at 4044 Å for K I and 6707 Å for Li I. Figure 1 displays the spectra, after removal of the undulating stellar continuum through the use of a low order polynomial.

Generally there are multiple interstellar clouds on any given line of sight whose existence is revealed by absorption at distinct Doppler velocities. The velocity structure has to be well known before an accurate lithium isotope ratio can be deduced from the spectra [9,10].

Therefore, the observed directions should have relatively simple velocity structure (i.e., one or two interstellar clouds with detectable Li I lines). In addition, data on another species, which has similar properties to lithium and resides in the same portion of the interstellar cloud, should be obtained for use as a template of the velocity structure. Our method differs slightly from the technique of Lemoine et al. [Ref. 9, 10]. Lemoine et al. used K I $\lambda 7699$ as their velocity template. This line of K I is much too strong to use as an effective velocity template. Instead, we obtained data on K I $\lambda 4044$ which is of comparable strength to the lithium lines (see Fig. 1) and therefore a more appropriate template. (The oscillator strength for $\lambda 4044$ is about a factor of 50 smaller [11].) Moreover, high signal to noise, high resolution spectra of the Li doublet at 6707 Å are required, since the lines are weak and the fine structure splitting for $^7$Li I is comparable to the isotope shift of $^6$Li I ($\sim 0.160$ Å), resulting in a blend of the $^7$Li and $^6$Li lines as shown in Fig. 1.

The observations were fitted by synthetic profiles where profile parameters were adjusted until a minimum $\chi^2$ was obtained; this method was successfully applied in our determination of the interstellar $^{11}$B/$^{10}$B ratio [12,13]. The results from applying the parameters obtained independently from the $^7$Li I and K I lines show agreement in column densities at the 1-$\sigma$ level. The results of profile fitting are displayed in Table 1, and Table 2 contains the inferred isotope ratios.

Our $^7$Li/$^6$Li ratio of $10.6 \pm 2.9$ for the cloud toward $\zeta$ Per is just consistent with a previous determination [16] ($5.5^{+1.3}_{-1.1}$), the difference being in the derived $N(^6$Li I$)$, and agrees with the most recent meteoritic value [7] (11.86-12.44). As shown by inspection of the line profile, the $^7$Li/$^6$Li ratios for o Per's clouds are highly non-solar: $^7$Li/$^6$Li $= 3.6 \pm 0.6$ and $1.7 \pm 0.3$, for the stronger and weaker cloud, respectively, with the latter being essentially the value ($\simeq 2$) predicted for Li production by cosmic ray-induced spallation. The most recent models [4-5] use up-to-date cross sections for spallation reactions and cosmic ray data. Reactions of importance for the Li isotopes are protons and helium nuclei on C, N, and O nuclei, and there is a contribution from helium-helium fusion. Considering recent results for the cosmic ray flux, composition, and propagation and incorporating the effects of ionization of interstellar gas, the computed $^7$Li/$^6$Li ratios are 1.5–1.7 [4] and $1.35 \pm 0.40$ [5], values equal to the lower of the ratios found for o Per. Our conclusion is that the cloud material has been subjected to a severe flux of energetic particles. Such a flux has not affected $\zeta$ Per's cloud. The line of sight toward o Per lies closer than that of $\zeta$ Per to a site of star formation, IC 348, where massive stars could supply energetic particles via supernovae or stellar winds. Confirmation is provided by a study of the interstellar OH and HD chemistry of these sight lines that showed the cosmic ray flux toward o Per was approximately an order of magnitude higher [17,18].

An earlier study [10] suggested a low $^7$Li/$^6$Li ratio for a cloud toward $\zeta$ Oph. Using the stronger K I line at 7699 Å as a velocity template, isotope ratios of $8.8 \pm 0.8 \pm 1.4$ and $1.4^{+1.2}_{-0.5} \pm 0.6$, where the second uncertainty represents systematic effects, were obtained. One concern with this two-component model for $\zeta$ Oph is that the main velocity component is known to be 2 components separated by 1 km s$^{-1}$ [19]. Another concern is that there is no evidence for molecular absorption from the second component toward $\zeta$ Oph, unlike the situation in other directions [19]. Because this earlier study probably used an inappropriate template, the difference from the Solar System ratio of 12.3 is at best suggestive. Our measurements, acquired at higher spectral resolution and analyzed with a more appropriate template, offer the first clear and unambiguous evidence for freshly synthesized products of cosmic ray spallation.

Derivation of the total interstellar abundance is not trivial because knowledge of the abundance for Li II, the dominant ion, and mechanisms for depletion onto interstellar grains is necessary. An estimate for Li II abundance from data on Li I requires the electron density, which can be approximated by the amount of C II, the most abundant element providing electrons in neutral interstellar gas. The resulting total Li/H abundances appear in Table 2. Since a Li isotope ratio of $\sim 2$ toward o Per suggests newly processed Li, the comparable Li/H abundances are unexpected. The two abundances, however, are very similar to recent interstellar Li results, using the analysis described in the legend to Table 2 for consistency: $13.1 \times 10^{-10}$ toward $\zeta$ Per [16], $11.0 \times 10^{-10}$ toward $\rho$ Oph [9], $20.6-24.4 \times 10^{-10}$ toward $\zeta$ Oph [16,10]. For further comparison, the Solar System value [6] is $20 \times 10^{-10}$. Since a weighted interstellar average of C II was utilized in determining $n_e$ toward o Per, there is additional uncertainty in the derived lithium abundance. There also remains the poorly known correction for depletion onto grains.

Another observational measure may shed light on this inconsistency, the K/Li ratio. It is not dependent on electron density or the strength of the radiation field, and the amounts of depletion onto grains are similar [25]. The inferred K/Li ratios displayed in Table 2 for the clouds toward o Per and $\zeta$ Per show no clear trend and are similar to the Solar System ratio [6] of $66 \pm 8$. Again, no indication for fresh Li is evident.

Although the measurements are as yet few in number, a reasonable assumption based on isotopic Li measurements for several clouds, local stars [26] and the Solar System is that $^7$Li/$^6$Li $\simeq$ 10 once characterized o Per's clouds. If gas with such an isotopic ratio is exposed to energetic particles, fresh $^7$Li and $^6$Li are synthesized but little lithium is destroyed. Dilution of $^7$Li in this fashion would necessarily not attain the observed low isotopic ratio until the total lithium abundance had been raised about an order of magnitude. Neither the Li abundance nor the K/Li ratio in o Per's clouds are, however, unusual; Li appears to be less abundant. This suggests that the energetic particles operated on gas highly depleted in lithium: ejecta from massive stars - winds and supernovae - may be the playground for the energetic particles.

Models [27] of superbubbles, which contain ionized gas blown by supernovae in star forming regions, show that energetic particles synthesize about $10^{51}$ Li atoms. At the end of the superbubble's existence, these Li atoms are diluted with swept up ambient material at which point the Li/H ratio is 100 times less than observed toward o Per. The clouds toward o Per, if spherical, contain $3 \times 10^{48}$ Li atoms in 3 solar masses of gas. If the superbubble is itself roughly spherical, the total Li content inferred from the o Per sight line is close to the predicted $10^{51}$ atoms. The superbubble scenario may account for our result provided that dilution of its products with ambient gas has not proceeded to completion. Sight lines containing partially mixed products from a superbubble like that toward o Per may be rare.

The presence of deuterium as HD molecules toward o Per [28] implies that the gas within the superbubble cannot have been totally depleted in Li initially. The D-containing gas maybe largely in the cloud for which we find $^7$Li/$^6$Li $\simeq$ 4. Synthesis of D within a superbubble is negligible. Existence of D may warrant consideration of alternative hypotheses, including

(i) isotopic fractionation of Li in diffuse clouds and (ii) the idea that pristine interstellar gas has $^7$Li/$^6$Li $\simeq 2$, the value for cosmic ray spallation, and varying degrees of contamination with ejecta from Li-rich red giants drive the ratio to the higher values seen in $\zeta$ Per's clouds and elsewhere.

# References

1. Reeves, H., Fowler, W.A., and Hoyle, F. Galactic cosmic ray origin of Li, Be, and B in stars. *Nature* **226**, 727–729 (1970).

2. Meneguzzi, M., Audouze, J., and Reeves, H. The production of the elements Li, Be, B by galactic cosmic rays in space and its relation with stellar observations. *Astr. Astrophys.* **15**, 337–359 (1971).

3. Meneguzzi, M., and Reeves, H. Light element production by cosmic rays. *Astr. Astrophys.* **40**, 99–110 (1975).

4. Ramaty, R., Kozlovsky, B., and Lingenfelter, R.E. Light isotopes, extinct radioisotopes, and gamma-ray lines from low-energy cosmic-ray interactions. *Astrophys. J.* **456**, 525–540 (1996).

5. Lemoine, M., Vangioni-Flam, E., and Cassé, M. Galactic cosmic rays and the evolution of light elements. *Astrophys. J.* **499**, 735–745 (1998).

6. Anders, E., and Grevesse, N. Abundances of the elements: meteoritic and solar. *Geochim. cosmochim. Acta* **53**, 197–214 (1989).

7. Chaussidon, M., and Robert, F. $^7$Li/$^6$Li and $^{11}$B/$^{10}$B variations in chondrules from the Semarkona unequilibrated chondrite. *Earth Planet. Sci. Lett.* **164**, 577–589 (1998).

8. Tull, R. G., MacQueen, P. J., Sneden, C., and Lambert, D. L. The high resolution cross-dispersed echelle white-pupil spectrometer of the McDonald Observatory 2.7 m telescope. *Publ. Astron. Soc. Pac.* **107**, 251–264 (1995).

9. Lemoine, M., Ferlet, R., Vidal-Madjar, A., Emerich, C., and Bertin, P. Interstellar lithium and the $^7$Li/$^6$Li ratio toward $\rho$ Oph. *Astr. Astrophys.* **269**, 469–476 (1993).

10. Lemoine, M., Ferlet, R., and Vidal-Madjar, A. The interstellar $^7$Li/$^6$Li ratio. The line of sight to $\zeta$ Ophiuchi. *Astr. Astrophys.* **298**, 879–893 (1995).

11. Morton, D.C. Atomic data for resonance absorption lines. I. Wavelengths longward of the Lyman limit. *Astrophys. J. Suppl.* **77**, 119–202 (1991).

12. Federman, S. R., Lambert, D. L., Cardelli, J. A., and Sheffer, Y. The boron isotope ratio in the interstellar medium. *Nature* **381**, 764–766 (1995).

13. Lambert, D. L., *et al.* The $^{11}$B/$^{10}$B ratio of local interstellar diffuse clouds. *Astrophys. J.* **494**, 614–622 (1998).

14. Sansonetti, C. J., Richou, B., Engleman, Jr., R., and Radziemski, L., J. Measurements of the resonance lines of $^6$Li and $^7$Li by Doppler-free frequency-modulation spectroscopy. *Phys. Rev.* **A52**, 2682–2688 (1995).

15. Hobbs, L.M., Thorburn, J.A., and Rebull, L.M. Lithium isotope ratios in halo stars. III. *Astrophys. J.* **523**, 797–804 (1999).

16. Meyer, D. M., Hawkins, I., and Wright, E. L. The interstellar $^7$Li/$^6$Li isotope ratio toward $\zeta$ Ophiuchi and $\zeta$ Persei. *Astrophys. J.* **409**, L61–L64 (1993).

17. van Dishoeck, E.F., and Black, J.H. Comprehensive models of diffuse interstellar clouds: Physical conditions and molecular abundances. *Astrophys. J. Suppl.* **62**, 109–145 (1986).

18. Federman, S. R., Weber, J., and Lambert, D. L. Cosmic ray-induced chemistry toward Perseus OB2. *Astrophys. J.* **463**, 181–190 (1996).

19. Crane, P., Lambert, D.L., and Sheffer, Y. A very high resolution survey of interstellar CH and CH$^+$. *Astrophys. J. Suppl.* **99**, 107–120 (1995).

20. Sofia, U. J., Cardelli, J. A., Guerin, K. P., and Meyer, D. M. Carbon in the diffuse interstellar medium. *Astrophys. J.* **482**, L105–L108 (1997).

21. Cardelli, J. A., Meyer, D. M., Jura, M., and Savage, B. D. The abundance of interstellar carbon. *Astrophys. J.* **467**, 334–340 (1996).

22. Savage, B.D., Bohlin, R.C., Drake, J.F., and Budich, W. A survey of interstellar molecular hydrogen. *Astrophys. J.* **216**, 291–307 (1977).

23. Diplas, A., and Savage, B.D. An *IUE* survey of interstellar H I Ly$\alpha$ absorption. I. Column densities. *Astrophys. J. Suppl.* **93**, 211–228 (1994).

24. Federman, S. R., *et al.* Chemical transitions for interstellar $C_2$ and CN in cloud envelopes. *Astrophys. J.* **424**, 772–792 (1994).

25. White, R. E. Interstellar lithium: differential depletion in diffuse clouds. *Astrophys. J.* **307**, 777–786 (1986).

26. Andersen, J., Gustafsson, B., and Lambert, D.L. The lithium isotope ratio in F and G stars. *Astr. Astrophys.* **136**, 65–73 (1984).

27. Parizot, E., and Drury, L. Superbubbles as the source of $^6$Li, Be, and B in the early Galaxy. *Astr. Astrophys.* **349**, 673–684 (1999).

28. Snow, T.P., Jr. Interstellar molecular abundances toward omicron Persei. *Astrophys. J.* **201**, L21–L24 (1975).

Acknowledgments: We thank E. Parizot and H. Reeves for fruitful discussions. This work was supported by grants from the National Aeronautics and Space Administration and made use of the Simbad database, operated at CDS, Strasbourg, France.

Table 1. **Results of Profile Fitting.** The synthesis included hyperfine structure for both $^6$Li and $^7$Li [14]. The hyperfine structure splitting of K I $\lambda$4044 was found to be negligible ($\sim$ 0.002 mÅ) and was not included in the analysis. The weaker member of the K I doublet is at 4047 Å and thus is not part of the analysis. For oscillator strengths, we adopted tabulated results for each hyperfine transition in Li I [15] and for the fine structure line of K I [11]. The profiles of the bluer $^7$Li I fine structure line and K I line were fitted first to obtain the $b$-value and velocity of the line in the local standard of rest ($v_{lsr}$), as well as column density ($N$) for each species. The $b$-value, which is dominated by turbulent motion, and velocity were then fixed and used in the fit of the Li I profile. The uncertainties in $N$ were determined from the uncertainties associated with the amount of absorption, while for the $b$-value and $v_{lsr}$, the results for the two fits provide a measure of their uncertainties. The final results toward each star are weighted averages determined by the separate fits. The results for component 1 toward o Per are listed above those for component 2. The larger $b$-value for the $^7$Li fit toward $\zeta$ Per arises in part from the November data having a slightly larger instrumental width. For the weak lines examined here, the difference in $b$-value produces a 10% effect in column density and in the inferred isotope ratio.

| Parameter | o Per | | | ζ Per | | |
|---|---|---|---|---|---|---|
| | $^7$Li fit | K fit | average | $^7$Li fit | K fit | average |
| $b$-value (km s$^{-1}$) | 2.1 | 2.1 | 2.1 | 1.9 | 1.4 | 1.7 |
| | 0.6 | 0.8 | 0.7 | | | |
| $v_{lsr}$ (km s$^{-1}$) | 6.5 | 6.5 | 6.5 | 6.5 | 6.5 | 6.5 |
| | 3.5 | 3.3 | 3.4 | | | |
| $N(^7$Li I$)$ ($10^8$ cm$^{-2}$) | 26 ± 2 | 26 ± 2 | 26 ± 1 | 37 ± 2 | 33 ± 2 | 35 ± 1 |
| | 5.6 ± 0.6 | 5.3 ± 0.8 | 5.5 ± 0.5 | | | |
| $N(^6$Li I$)$ ($10^8$ cm$^{-2}$) | 7.0 ± 1.5 | 7.4 ± 1.5 | 7.2 ± 1.1 | 3.7 ± 1.4 | 3.0 ± 1.1 | 3.3 ± 0.9 |
| | 3.2 ± 0.6 | 3.2 ± 0.8 | 3.2 ± 0.5 | | | |
| $N($K I$)$ ($10^{11}$ cm$^{-2}$) | 9.4 ± 0.3 | 9.4 ± 0.4 | 9.4 ± 0.2 | 7.6 ± 0.4 | 7.0 ± 0.4 | 7.3 ± 0.3 |
| | 1.8 ± 0.2 | 1.9 ± 0.2 | 1.9 ± 0.1 | | | |

Table 2. **Results for the $^7$Li/$^6$Li Ratio, the Total Li/H Abundance and the K/Li Abundance Ratio.** For our estimates of the total Li/H abundance, the electron density ($n_e$) is obtained from $N$(C II), the total proton column density $[N(\text{H}) = N(\text{H I}) + 2N(\text{H}_2)]$, and the gas density ($n$). Since no precise $N$(C II) measurements exist for the direction toward o Per, the weighted mean interstellar ratio [20] of $N(\text{C II})/N(\text{H}) = (1.42 \pm 0.13) \times 10^{-4}$ was utilized, yielding $N(\text{C II}) = (2.2 \pm 0.8) \times 10^{17}$ cm$^{-2}$. The value of $N$(C II) is $(1.84 \pm 0.32) \times 10^{17}$ cm$^{-2}$ toward $\zeta$ Per [21]. The columns $N$(H) [22,23] are $(15.2 \pm 5.3) \times 10^{20}$ cm$^{-2}$ toward o Per and $(15.8 \pm 4.7) \times 10^{20}$ cm$^{-2}$ toward $\zeta$ Per. Chemical models for carbon-bearing molecules [24] indicate values for $n$ of 800 and 700 cm$^{-3}$ toward o and $\zeta$ Per, respectively. The total Li/H abundance is derived from $\{[N(^7\text{Li I}) + N(^6\text{Li I})]/N(\text{H})\} \times [G/(\alpha n_e)]$, where a value of 41 was used for $(G/\alpha)_{Li}$, the photoionization rate to recombination rate coefficient [25]. The ionization corrections for the two clouds toward o Per are assumed to be the same. The K/Li ratios are based on column densities in Table 1 and a value of 9.4 for $(G/\alpha)_K$ [25]. For the $^7$Li/$^6$Li and K/Li ratios toward o Per, component 1 is the first entry and component 2 is the second.

| Parameter | o Per | $\zeta$ Per |
|---|---|---|
| $^7$Li/$^6$Li ratio | $3.6 \pm 0.6$ | $10.6 \pm 2.9$ |
|  | $1.7 \pm 0.3$ |  |
| Li/H abundance | $(9.8 \pm 3.5) \times 10^{-10}$ | $(12.2 \pm 2.2) \times 10^{-10}$ |
| K/Li abundance ratio | $65 \pm 4$ | $42 \pm 2$ |
|  | $48 \pm 4$ |  |

**Figure Caption**

Fig. 1. Interstellar spectra of K I (a and c) and Li I (b and d) toward $\zeta$ and o Per. The spectra were reduced with NOAO SUN/IRAF. After correcting for the bias and removing scattered light, the stellar images were flat fielded to remove pixel-to-pixel variations in sensitivity. Extracted spectra were produced by summing over pixels perpendicular to the dispersion direction and were placed on a wavelength scale with spectra from a Th-Ar hollow cathode taken throughout the night. Individual spectra were Doppler corrected and coadded, yielding a signal-to-noise ratio per pixel of approximately 2500:1 for each stellar spectrum based on the peak-to-peak variation in the continuum. The data points are filled circles. The dot-dashed line is the fit to the line profiles using the K I line for the velocity structure. The solid line gives the difference between data and fit; the offset is 1.002. The vertical lines in the K I plot for o Per show the positions of the two velocity components; a single velocity component is seen toward $\zeta$ Per. The vertical dashes in the Li I plot for $\zeta$ Per indicate positions of the $^7$Li and $^6$Li fine structure lines. In the absence of $^6$Li, an unsaturated $^7$Li feature from a single cloud is a doublet with the red component half the strength of the blue component. A $^6$Li contribution augments the strength of the red component and contributes a weaker redder component. As long as the $^7$Li/$^6$Li ratio is not small, the Li I feature is dominated by the 2:1 ratio of the blue to red component of (primarily) $^7$Li. For o Per, however, the red component approaches the strength of the blue component. Since K I $\lambda$4044 shows that this enhancement cannot be due to a red-shifted $^7$Li component, the clouds toward o Per must be rich in $^6$Li.

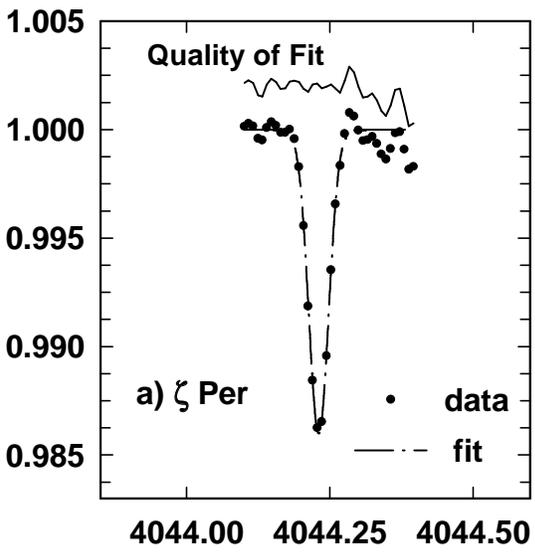
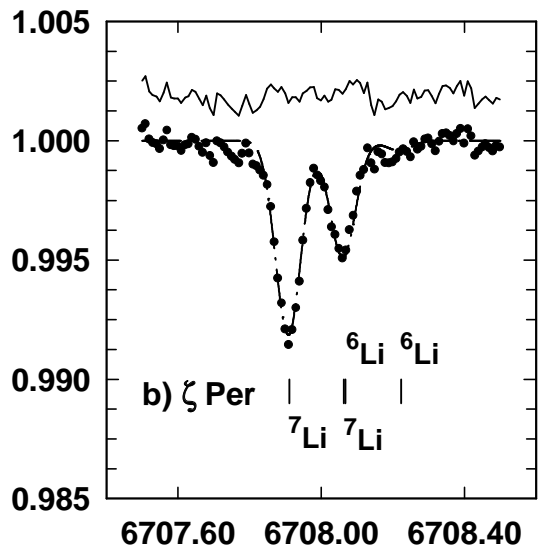
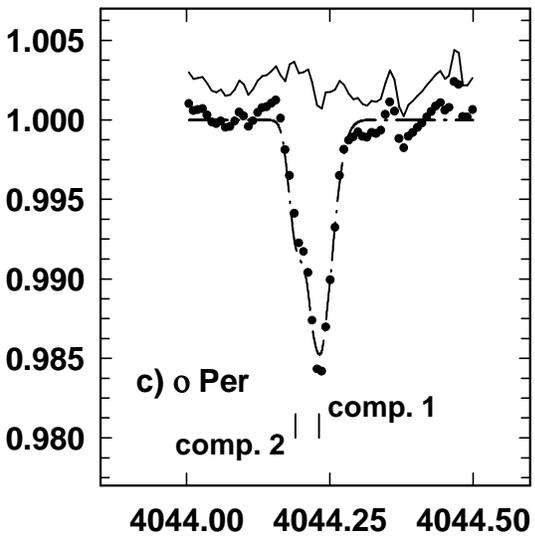
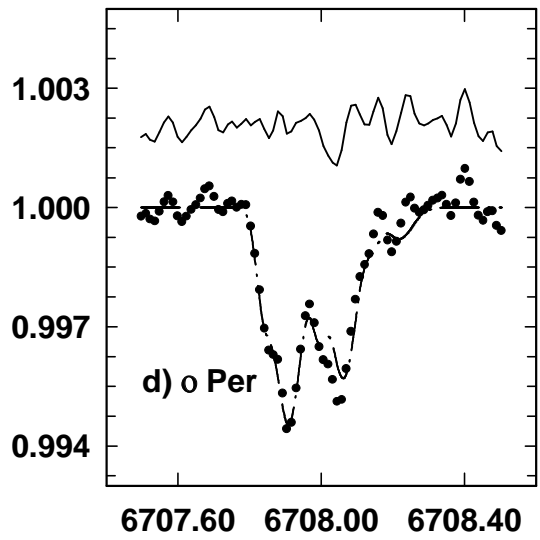